%% file: GMP12_PRL.tex
\documentclass[notitlepage,prd,twocolumn,showpacs,preprintnumbers,floatfix,nofootinbib,superscriptaddress]{revtex4-2}
\usepackage{amssymb,amsmath,placeins,epsfig,array}
\usepackage[usenames,dvipsnames]{color}
\usepackage[normalem]{ulem}
\usepackage[breaklinks,colorlinks,urlcolor=blue,citecolor=blue,linkcolor=blue]{hyperref}
\usepackage{lineno}
\usepackage{float}
\usepackage{mathptmx}
\include{commands}

\begin{document}
\title{Form factors and two-photon exchange in high-energy elastic electron-proton scattering} 

\include{collaboration}

\begin{abstract}

We present new precision measurements of the elastic electron-proton scattering cross section for momentum transfer (Q$^2$) up to 15.75~\gevsq.
Combined with existing data, these provide an improved extraction of the proton magnetic form factor at high Q$^2$ and double the range over which a longitudinal/transverse separation of the cross section can be performed.
The difference between our results and polarization data agrees with that observed at lower Q$^2$ and attributed to hard two-photon exchange (TPE) effects, extending to 8~(GeV/c)$^2$ the range of Q$^2$ for which a discrepancy is established at $>$95\% confidence. We use the discrepancy to quantify the size of TPE contributions needed to explain the cross section at high Q$^2$.

\end{abstract}

\date{\today}

\maketitle


Elastic electron scattering is a key process used in studies of matter across a wide range of energy scales and in many subfields of physics. In the one-photon exchange approximation (OPE), first calculated in Ref.~\cite{Rosenbluth_1950}, the differential electron-nucleon elastic scattering cross section, $d\sigma (\theta_e)/d\Omega_e$, is the product of the cross section for a structureless object and a structure-dependent term that depends on the Sachs magnetic and electric form factors~\cite{Sachs:1962zzc}, G$_{_M}($Q$^2)$ and G$_{_E}($Q$^2)$, which encode the spatial distributions of magnetization and charge in the proton:
\begin{equation}
\frac{d\sigma(\theta_e)}{d\Omega_e} = \frac{d\sigma_{_{Mott}}}{d\Omega_e} \cdot \frac{\tau \, {\rm G}_{_M}^2({\rm Q}^2) + \varepsilon \, {\rm G}_{_E}^2({\rm Q}^2)}{ \varepsilon (1+\tau)}.
\label{eq:rosenbluth}
\end{equation}
In Eq.~\eqref{eq:rosenbluth}, $\theta_e$ is the scattering angle of the electron, ${d\sigma_{_{Mott}}}/{d\Omega_e}$ is the cross section for scattering of an electron  with incident (scattered) energy $E_e$ ($E'_e$) from a structureless target, $Q^2 = 4 E_e E_e' \sin^2(\theta_e/2)$ is the negative four-momentum transfer squared, $\varepsilon \equiv \left[1 + 2(1+\tau) \tan^2(\theta_e/2)\right]^{-1}$ is the virtual photon polarization parameter, and $\tau \equiv Q^2/4M_p^2$.
The structure-dependent term is isolated in the reduced cross section, 
\begin{eqnarray}
\sigma_{_R} &=& \tau \ {\rm G}_{_M}^2({\rm Q}^2) + \varepsilon \ {\rm G}_{_E}^2({\rm Q}^2) = \sigma_{_T} + \varepsilon \ \sigma_{_L} \nonumber \\
            &=& {\rm G}_{_M}^2({\rm Q}^2) (\tau + \varepsilon \ {\rm RS(Q}^2)/\mu_p^2),
\label{eq:RS}
\end{eqnarray}
where $\sigma_{_L}$ and $\sigma_{_T}$ are the longitudinal and transverse contributions to the cross section, respectively, RS $=(\mu_p$\gep/\gmp)$^2$ is the normalized Rosenbluth slope, and $\mu_p$ is the proton magnetic moment.
The form factors can be extracted using measurements at fixed \qsq~but different values of $\varepsilon$, corresponding to different electron scattering angles.
A linear fit to measurements of $\sigma_{_R}(\varepsilon)$ yields an intercept of $\sigma_{_R}(\varepsilon=0) = \tau \, {\rm G}_{_M}^2$, and a slope of $d\sigma_{_R}/d\varepsilon = {\rm G}_{_E}^2$. This method is commonly known as Rosenbluth or Longitudinal/Transverse (L/T) separation.

Pioneering measurements of elastic electron-proton scattering by R.~Hofstadter~\cite{Hofstadter:1956qs} confirmed the theoretical expectation of linear dependence of $\sigma_{_R}$ as a function of $\varepsilon$, which supported the use of the OPE approximation.
Theoretical studies of the effects beyond OPE were performed soon after that~\cite{Drell-1957, Drell-1959}.
Early experimental searches using recoil proton polarization~\cite{Bizot-1965} and lepton charge asymmetry~\cite{Mar-1968, Bouquet-1968} failed to find significant deviations from the OPE approximation.
A number of precision measurements, including one in Ref.~\cite{Walker:1993vj}, extended linearity tests up to \qsq $=$ 3~\gevsq~and demonstrated that \gep~and \gmp~both approximately follow the dipole form $G_{_D} \equiv \left(1+\text{Q}^2/\Lambda^2\right)^{-2}$, with $\Lambda^2 = 0.71$ GeV$^2$, yielding form factor scaling: $\mu_p$\gep/\gmp $\approx 1$.
At larger \qsq~values, $\tau$ enhances the contribution from G$_{_M}^2$ to the cross section, making it difficult to extract G$_{_E}^2$.
Analyses at higher \qsq~values~\cite{Kirk:1972xm, Sill:1992qw} extracted \gmp~under the assumption that RS = 1, and found that
Q$^4 \, $G$_{_M}$(\qsq) was \qsq-independent above 10~\gevsq, consistent with pQCD predictions~\cite{Lepage:1979za, *Lepage:1980fj}. 

The reduced sensitivity to \gepsq~at high \qsq~in the Rosenbluth method motivated the use of double polarization observables~\cite{Akhiezer:1957aa}, for which the OPE formalism was developed in Refs.~\cite{*Akhiezer:1968ek,*Akhiezer:1974em,Frolov:1958aa,Scofield:1959zz,*Scofield:1966zz,Arnold:1980zj}. 
Polarization measurements are directly sensitive to the ratio ~\gep/\gmp, but not to the individual form factors. 
About 20 years ago the first precision measurements of \gep/\gmp~for Q$^2$ up to several \gevsq~were performed using the  polarization transfer method~\cite{Jones:1999rz} and a novel effect was discovered: the form factor ratio (FFR) extracted from polarization data decreased dramatically with \qsq~\cite{Jones:1999rz,*Gayou:2001qd, *Punjabi:2005wq, Puckett:2011xg, Puckett:2010ac,*Puckett:2017flj}.

The decrease of the FFR with increasing \qsq, an unexpected effect, implied a significant reduction of \gep, with theoretical 
explanations ranging from the role of quark orbital momentum~\cite{Miller:2002qb,*Belitsky:2002kj} to the effect of the diquark correlation in the nucleon ground state~\cite{Roberts:2007jh, *Cloet:2008re, *Barabanov:2020jvn}.
In addition, the difference between the FFR extracted from the polarization measurements and from the cross section results was surprising, and requires deeper understanding.
This difference is referred to henceforth as the form factor ratio puzzle (FFRP).

A reanalysis of the world data on RS~\cite{Arrington:2003df, *Arrington:2003qk}, and new measurements of RS values with both scattered electron detection~\cite{Christy:2004rc} and recoil proton detection~\cite{Qattan:2004ht}, confirmed with improved precision the original observation of form factor scaling, enhancing the FFRP. 
Assuming that there are no unexpected
errors with the now extensive body of Rosenbluth and polarization measurements, and that the radiative corrections (RC) applied are complete (except for the excluded hard TPE contributions), the only remaining explanation within the Standard Model is two-photon-exchange (TPE) or higher-order corrections.
The hard TPE contributions are defined in this context as the TPE terms omitted in conventional radiative correction procedures which include only the IR-divergent terms, meaning that the definition of hard TPE depends slightly on the RC prescription~\cite{Maximon:2000hm}.

An analysis of world data found that non-linearities in the reduced cross section as a function of $\varepsilon$, indicating deviations from the OPE picture, were extremely small~\cite{Tvaskis:2005ex}, although the lack of non-linear contributions does not rule out a change to the slope that could explain the FFRP.
At large Q$^2$ values, where the slope arising from $G_E^2$ is small, even a tiny change of RS can modify the extraction of \gep/\gmp~significantly. 

TPE cross section contributions have the opposite sign for electron and positron scattering, making a comparison of $e^+$-$p$ and $e^-$-$p$ scattering one of the most direct tests for TPE. 
A global re-examination of electron/positron scattering comparisons in 2003 showed evidence for TPE~\cite{Arrington:2003ck} at low \qsq~values. After 2010, new experiments were performed to improve the precision and extend the kinematic range of these comparisons~\cite{Rachek:2014fam, Adikaram:2014ykv,*Rimal:2016toz,Henderson:2016dea}, observing clear hard TPE in the ratio of $e^+$-$p$ and $e^-$-$p$ elastic scattering up to Q$^2 \approx 2$~\gevsq. 
Finally, the contribution of TPE to polarization transfer observables was found to be small~\cite{PhysRevLett.106.132501}, as predicted by calculations~\cite{Borisyuk:2014ssa, Blunden:2005ew, Afanasev:2005mp}.
Given this empirical understanding, the discrepancy (FFRP) is 
taken in our study as a measure of the TPE impact on the cross section, as in Refs.~\cite{Arrington:2003df, Arrington:2003qk, Qattan:2011ke, A1:2013fsc}.

While most examinations of the FFRP focus on hard TPE, any $\varepsilon$-dependent correction would contribute to the discrepancy, leading to new examinations of the full radiative correction procedures~\cite{Maximon:2000hm, Gerasimov:2015aoa, Gramolin:2016hjt}.
The most recent and complete update~\cite{Gramolin:2016hjt} was applied to SLAC data~\cite{Walker:1993vj, Andivahis:1994rq}, yielding a reduced discrepancy still providing a clear confirmation of the FFRP for \qsq~from 4-7~\gevsq.
Notwithstanding these experimental and theoretical efforts, a full calculation of the TPE contribution is still not available mostly due to its dependence on the hadron structure of the intermediate states (see reviews~\cite{Arrington:2011dn, Afanasev:2017gsk, Ye:2017gyb}).

This work provides new experimental data at very large \qsq, addresses the significance of the FFRP at much higher values of Q$^2$ than previously investigated, provides reanalyzed RC for six previous experiments, and improves the precision of experimental constraints on TPE effects in elastic $e$-$p$ scattering.
Our new low-$\varepsilon$ data, combined with existing high-$\varepsilon$ measurements~\cite{Kirk:1972xm, Rock:1991jy, Sill:1992qw, Andivahis:1994rq}, provide new Rosenbluth separations of \gep and \gmp~above 7~\gevsq ~and significantly improved precision in the extraction of \gmp.
Our new data also provide an important baseline for high-\qsq~measurements enabled by the 12~GeV upgrade at Jefferson Lab, where precise knowledge of the elastic cross section is needed for experimental normalization and cross checks, and as input to the broader program of high-\qsq~proton and neutron structure measurements.
 
This experiment, referred to hereafter as GMp12, was performed in Hall A of Jefferson Lab using the basic suite of experimental instrumentation~\cite{Alcorn:2004sb}. 
A 100\% duty-factor electron beam with current up to 68~$\mu$A and energy from 2.2 to 11 GeV was incident on a 15-cm long liquid hydrogen target.
The target operated at a temperature of 19~K, a pressure of 25 psia, and a density of 0.0732 g/cm$^3$.
The hydrogen target was complemented by a ``dummy" target consisting of two aluminum foils, used to measure and subtract events originating from the entrance and exit windows of the hydrogen cell. 
The target density reduction with increasing beam intensity, due to localized boiling of the cryogen, was found to
be 2.7\% per 100~$\mu$A,~\cite{Barak2018}, with an uncertainty in the variation across the current range of the experiment of $0.35\%$ 

The energy of incident electrons was determined using the Hall A ARC energy measurement system,~\cite{Berthot:1999jp}, which measures the field integral of the dipoles which bend the beam through 34.257 degrees from the accelerator into Hall A.  
These results were cross checked with spin precession studies and beam energy measurements in Hall C.
The uncertainty in the beam energy was found to be less than 0.1\% for all kinematics~\cite{Santiesteban2020aa}. 
The beam current was measured by beam charge monitors (BCMs)~\cite{Thir2019}, which were calibrated against a well-understood Unser monitor~\cite{Unser:1981fh}.
The uncertainty on the beam current and accumulated charge was defined by the accuracy of the BCM calibration.
An absolute uncertainty of 0.06~$\mu$A stems from the current source utilized to calibrate the Unser monitor.  
The latter results in an uncertainty of 0.1\% at a current of 65~$\mu$A, utilized for most of the GMp12 kinematics, and up to a maximum of 0.6\% for the lowest current of 10~$\mu$A.

The scattered electrons were detected in the left and right Hall A High Resolution Spectrometers (LHRS and RHRS, respectively), with the central momentum of the spectrometers set to detect elastically scattered electrons.
The HRSs have a solid angle acceptance of 6.0~msr, momentum acceptance of $\pm$4.5\%, intrinsic momentum resolution of $2.5 \times 10^{-4}$, and angular resolution of 0.6~mrad.
The primary trigger was formed as a coincidence of signals in the front and back scintillator planes (separated by two meters) and the gas Cherenkov counter.
The trigger efficiency was monitored using a sample of triggers that required only two of these three signals.
For this experiment, the tracking system in each HRS was upgraded by adding a three-layer straw tube drift chamber to allow accurate determination of the track reconstruction efficiency~\cite{Wang2017}.
The particle identification detectors included a two-layer shower detector and a gas Cherenkov counter with enhanced light collection efficiency by means of a wavelength shifter~\cite{Allada:2015kfa}.
Dead times of the trigger counters, front-end electronics, and DAQ were constantly measured using pulser generated events~\cite{Barak2018, Longwu2018}.
The uncorrelated systematic uncertainty of the GMp12 cross section data is 1.2-1.3\%, while the overall normalization uncertainty is 1.6\% (2.0\%) for the LHRS (RHRS) data. 
A more detailed breakdown and discussion of the main systematic uncertainties is presented in the supplemental material~\cite{supplemental}.

Full simulations of the incident electron-target interaction and the electron trajectory through the HRS magnets and detectors were performed for each kinematic setting using an updated version of the magnetic optics Monte Carlo code~\cite{Arrington1998} incorporating the HRSs. 
The event distributions in the detector package were compared with the simulated data and used to fine-tune the model of the HRS optical transport.  
Radiative processes were implemented using the approach built into the Monte Carlo simulation, described in Ref.~\cite{Ent:2001rc}, based on an updated implementation~\cite{Walker:1993vj} of the RC formalism of Ref.~\cite{Mo:1968cg}.
The resulting cross section values from GMp12 were then adjusted to account for the difference between the prescription above and the RC calculation of Refs.~\cite{Maximon:2000hm, Gerasimov:2015aoa, Gramolin:2016hjt} which has the most accurate evaluation of the internal and external radiation. This is essential for the analysis in the present work, as it resolves some of the discrepancy seen in past comparisons based on older radiative correction procedures.

The kinematics and reduced cross section results from the GMp12 experiment are shown in Table~\ref{tab:T1}.
\begin{table}[ht]
\caption{Kinematics and reduced cross sections for GMp12, with statistical and point-to-point systematic uncertainties added in quadrature. Points labeled with an asterisk (*) were taken with the RHRS. There is an additional 1.6\% (2.0\%) normalization uncertainty for the LHRS (RHRS) data.}
\label{tab:T1}
\centering
    \begin{tabular}{c|c|c|c|c}
\hline \hline
$E_e$    & $\theta_e$  & \qsq        & $\varepsilon$ & $\sigma_{_R}$ (Eq.~\ref{eq:RS}) \\ 
(GeV)    & (deg)  & (GeV/c)$^2$ &  &  \\ \hline

    2.222   	&~~42.001~~& 1.577 &~~0.701~~&~$(4.273 \pm 0.040)\times 10^{-2}$~ \\  
   ~2.222$^*$	&  48.666  & 1.858 &~~0.615~~& $(2.983 \pm 0.057)\times 10^{-2}$ \\  
    6.427		&  24.250  & 4.543 &~~0.826~~& $(3.813 \pm 0.057)\times 10^{-3}$ \\  
    6.427		&  30.909  & 5.947 &~~0.709~~& $(1.805 \pm 0.025)\times 10^{-3}$ \\ 
    6.427		&  37.008 & 6.993 &~~0.599~~& $(1.113 \pm 0.016)\times 10^{-3}$\\ 
    6.427		&  44.500 & 7.992 &~~0.478~~& $(7.289 \pm 0.109)\times 10^{-4}$\\ 
    8.518		&  30.909 & 9.002 &~~0.648~~& $(5.163 \pm 0.078)\times 10^{-4}$\\ 
   ~6.427$^*$	&  55.900 & 9.053 &~~0.332~~& $(4.859 \pm 0.107)\times 10^{-4}$\\ 
    8.518		&  34.400 & 9.807 &~~0.580~~& $(3.923 \pm 0.059)\times 10^{-4}$\\  
   ~8.518$^*$	&  42.001 & 11.19 &~~0.448~~& $(2.565 \pm 0.041)\times 10^{-4}$\\  
   ~8.518$^*$	&  48.666 & 12.07 &~~0.356~~& $(1.933 \pm 0.043)\times 10^{-4}$\\
   ~8.518$^*$	&  53.501 & 12.57 &~~0.301~~& $(1.664 \pm 0.053)\times 10^{-4}$\\ 
    10.587~  	&  48.666 & 15.76 &~~0.309~~& $(8.405 \pm 0.227)\times 10^{-5}$ \\ \hline \hline
    \end{tabular}
\end{table} 

We combine our results with cross sections from several JLab and SLAC experiments~\cite{Sill:1992qw, Walker:1993vj, Christy:2004rc, Kirk:1972xm, Rock:1991jy, Andivahis:1994rq} spanning a \qsq~range of 0.4-31 \gevsq~in a global fit of the \qsq~and $\varepsilon$ dependence of the elastic cross section using Eq.~\ref{eq:RS}.
These experiments, comprising 121 kinematic points, were chosen because the publications provide sufficient information on their RC procedures and cutoffs to allow us to self-consistently implement the RC modification~\cite{Gramolin:2016hjt}.
The normalizations of the data for the individual experiments were allowed to vary based on their quoted normalization uncertainties, except for the data of Ref.~\cite{Christy:2004rc}, which cover a wide range of Q$^2$ with the best accuracy.
The cross sections were fit in terms of \gmp~and RS with the following simple parametrization:
\begin{eqnarray}\label{eq:fit}
\nonumber
G_{_M} &=& \mu_p \, (1 \, +\, {a_1 \tau})/{(1} \, +\, b_1\tau \, +\, b_2\tau^2 \, +\, b_3\tau^3), \\
\mathrm {RS} &=&  1 \, + \, {c_1 \tau} + \, {c_2 \tau^2}.
\end{eqnarray}
The fit gives $\chi^2=88.7$ for 107 degrees of freedom; the parameters and uncertainties are given in Tab.~\ref{tab:fitsummary}. The cross section database and the full covariance matrix of the fit parameters are given in the supplemental material~\cite{supplemental}. 

\begin{table}[!htb]
\caption{Fit parameters and uncertainties (Eq.~\ref{eq:fit})}
  \begin{center}
      \begin{ruledtabular}
      \begin{tabular}{cccccc}
        $a_1$ & $b_1$ & $b_2$ & $b_3$ & $c_1$ & $c_2$ \\  \hline
         0.072(22) & 10.73(11) & 19.81(17) & 4.75(65) & -0.46(12) & 0.12(10)
      \end{tabular}
    \end{ruledtabular}
    \label{tab:fitsummary}
  \end{center}
\end{table}

\begin{figure}[htb]
\includegraphics[width=0.95\columnwidth, height=9.5cm, trim = {1mm 3mm 10mm 5mm}, clip]{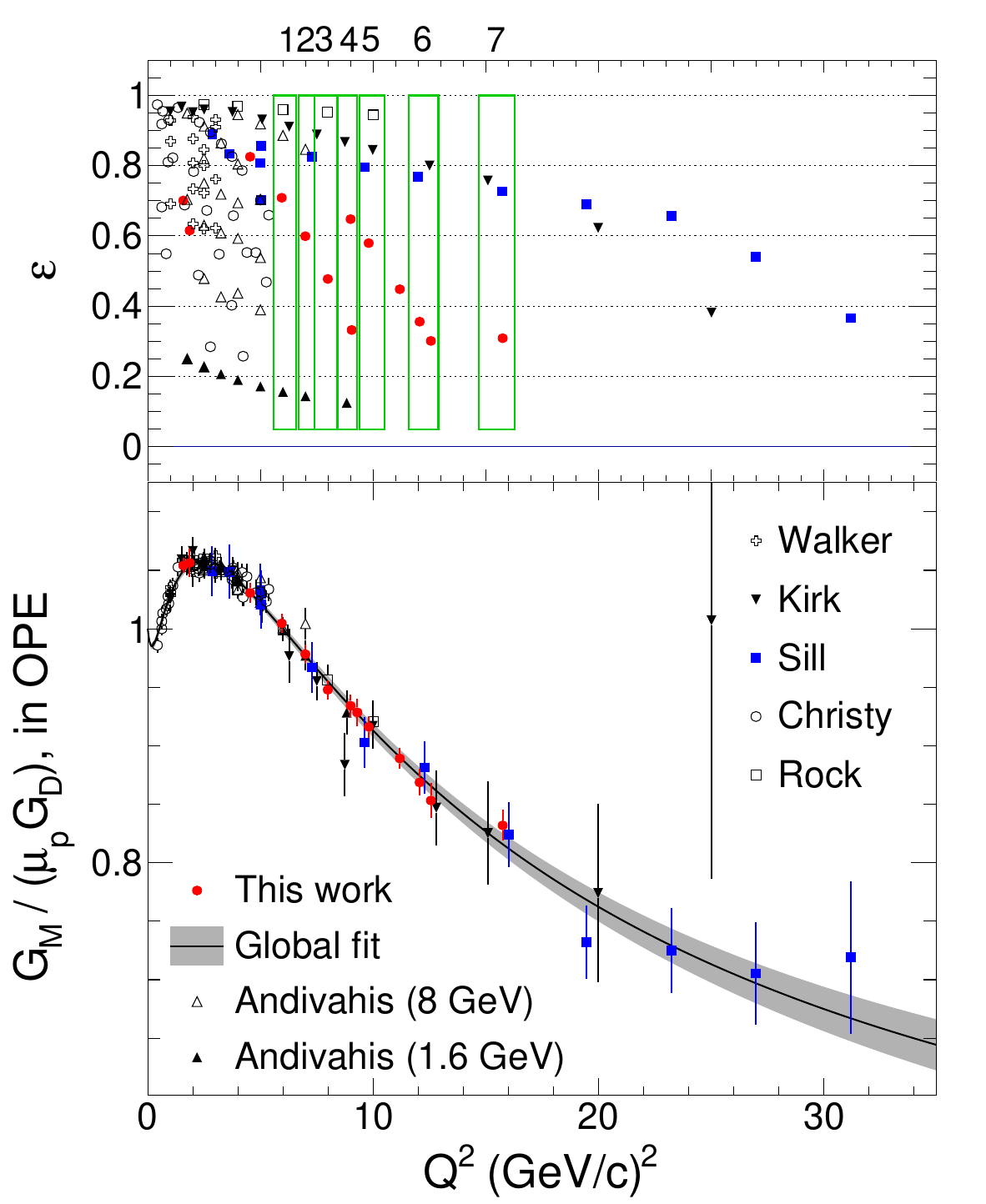} 
\caption{(Top) Kinematics of elastic $e$-$p$ data, Refs.~\cite{Sill:1992qw, Walker:1993vj, Christy:2004rc, Kirk:1972xm, Rock:1991jy, Andivahis:1994rq} and this work, used in the global fit and Rosenbluth separations; boxes (1-7) indicate the groupings of points for the Rosenbluth separations. 
(Bottom) Effective proton magnetic form factor, normalized by the standard dipole $\mu_p G_{_D}$, obtained from the cross section measurements. The curve shows the result of our global fit, with the gray shaded area indicating the 68\% confidence interval.}
\label{fig:GMp}
\end{figure}
Figure~\ref{fig:GMp} shows the global fit to \gmp~along with the values extracted from individual cross section measurements using the fit to RS(Q$^2$) to extrapolate to $\varepsilon=0$. Our new data reduce the high-Q$^2$ uncertainties on \gmp in the global fit by $>30\%$.

\begin{figure}[htb]
\includegraphics[width=0.95\columnwidth]{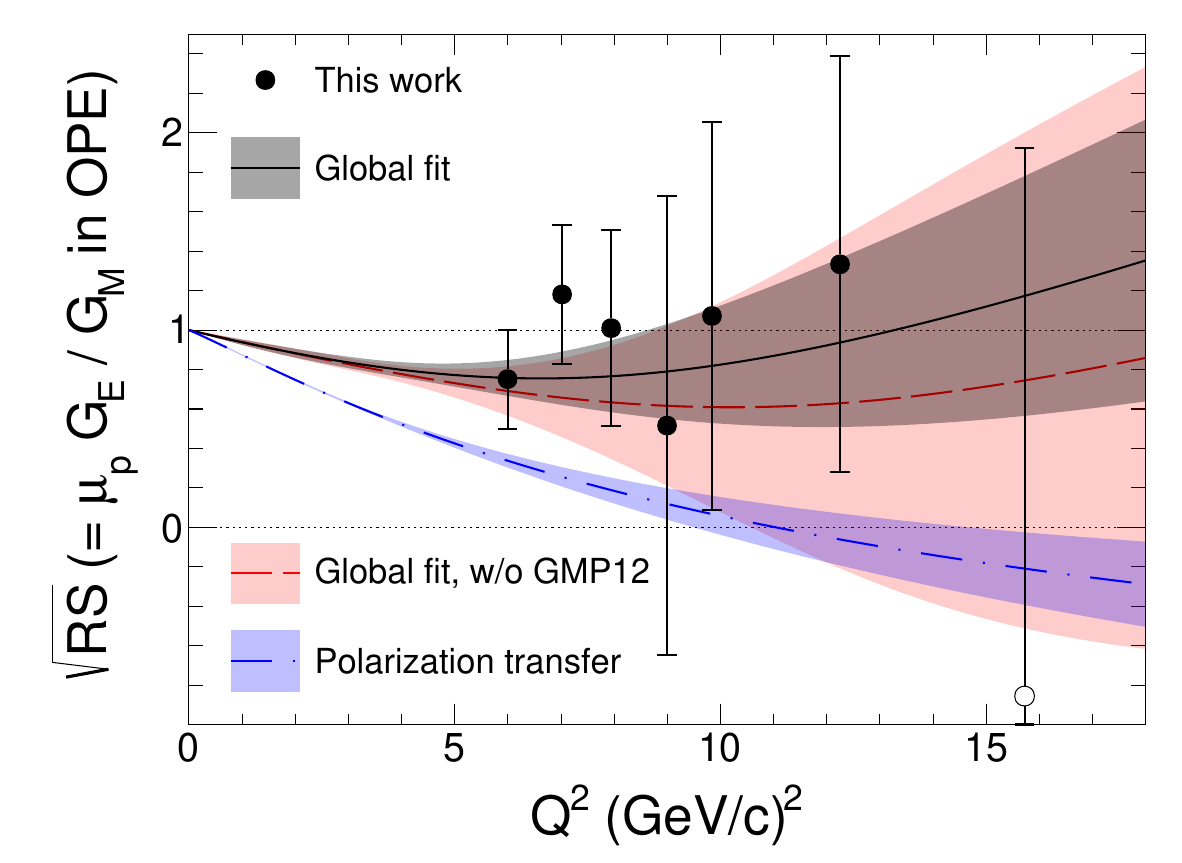}
	\caption{Direct Rosenbluth separation results
for $\sqrt{\text{RS}}$ ( $ = \mu_p$G$_{_E}$/G$_{_M}$ in OPE). The black solid (red dashed) curve shows the results of our fit to the cross section data with (without) the new GMp12 data. The blue dot-dashed curve shows $\mu_p$G$_{_E}$/G$_{_M}$ from a fit to the polarization data~\cite{supplemental}. The shaded bands show the 68\% confidence intervals of the respective fits. We plot $-\sqrt{\left|\text{RS}\right|}$ for the highest Q$^2$ point (an open circle), where RS $<$ 0.}  	
\label{fig:RS}
\end{figure}

We also performed direct Rosenbluth separations by grouping together points with similar \qsq~values, as indicated by the boxes in the top panel of Fig.~\ref{fig:GMp}. 
The normalization resulting from the global fit was applied to each data set, modifying the cross sections from Table~\ref{tab:T1}, and the data in each Q$^2$ bin were interpolated to a common Q$^2_c$ value using the global fit~\cite{supplemental}.
\gep and \gmp were then extracted from a linear fit to the $\varepsilon$ dependence of $\sigma_R$ for each of the seven Q$^2$ bins. The results of this extraction are given in Table~\ref{tab:T2}.
Figure~\ref{fig:RS} shows $\sqrt{\text{RS}}$ (yielding $\mu_p$G$_{_E}$/G$_{_M}$ in the OPE) from our global analysis, along with a fit to the polarization data.

While it is conventional to compare measurements by showing 
$\sqrt{\text{RS}} = \mu_p$G$_{_E}$/G$_{_M}$, it is more correct to use RS which is the observable most directly extracted from the cross sections. Our quantitative comparisons of the FFRP use RS, as detailed in Ref.~\cite{supplemental}.
We find that the cross section data, using the best available radiative corrections but excluding hard TPE contributions, show a 2$\sigma$ discrepancy with the polarization data up to 8~GeV$^2$ (1$\sigma$ up to 14~GeV$^2$).
 
Accommodating this discrepancy at large $Q^2$ values requires a TPE contribution that reduces the cross section by $\sim$4\% at $\varepsilon=0$, assuming a linear $\varepsilon$ dependence and a vanishing TPE contribution at $\varepsilon = 1$, as detailed in the supplemental material~\cite{supplemental},  which includes Refs.~\cite{Chen:2007ac, Kelly:2004hm, Makino:2006sx, Walker_thesis, Tsai:1961zz, Tsai_1971, Ron:2011rd, Zhan:2011ji, Paolone:2010qc, Pospischil:2001pp, Jones:2006kf, Crawford:2006rz, Milbrath:1997de, Gayou:2001qt, Github, CrossSectionDatabase, FitDetails}.
The cross section has a $\sim$2\% variation over the typical $\varepsilon$ range of the data.
This is qualitatively consistent with some high-Q$^2$ calculations~\cite{Blunden:2005ew, Afanasev:2017gsk} that predict large deviations from linear $\varepsilon$ dependence which, however, are most significant below the $\varepsilon$ range of the current data. 
Note that without the updated radiative corrections applied in this analysis, the discrepancy would have required TPE with a $\sim$6.5\% linear $\varepsilon$ dependence, consistent with previous estimates~\cite{Arrington:2003qk, Arrington:2007ux} based on analyses of data at lower Q$^2$ values using the older RC procedures.

In summary, the $e$-$p$ elastic scattering cross section was measured for beam energies in the range of 2.2 - 11~GeV and \qsq~up to 15.75~\gevsq.
These new, high-precision cross sections provide an important baseline for the future proton and neutron structure investigations in the Jefferson Lab 12 GeV program.
Our data were combined with existing cross section measurements~\cite{Walker:1993vj, Sill:1992qw, Andivahis:1994rq, Rock:1991jy, Kirk:1972xm} to perform Rosenbluth separations in a new \qsq~regime.
The observed difference between the measured Rosenbluth slope and the OPE expectation, based on \gep/\gmp~from polarization data, would be resolved with a $\sim$4\% contribution to the cross section from hard TPE up to Q$^2=8$~GeV$^2$, with no indication of significant Q$^2$ dependence at large Q$^2$ values. 

\begin{table}[htb]
\caption{Rosenbluth separation results for the data groupings shown in the top panel of Fig.~\ref{fig:GMp}, after centering to the average $Q^{2}_c$. The quoted values of $\sigma_{_L}$ and $\sigma_{_T}$ as defined in Eq.~\ref{eq:RS}, and G$_{_M}/(\mu_p\text{G}_{_D})$ and $\mu_p$G$_{_E}$/G$_{_M}$ are obtained assuming validity of the OPE approximation. For the largest Q$^2$, where $\sigma_{_L}<0$, we quote $-\sqrt{\left|\text{RS}\right|}$.} 
\label{tab:T2}
\centering
\begin{ruledtabular}
\begin{tabular}{c|c|c|c|c}
Q$^{2}_c$       & $\sigma_{_T}\times 10^5$	&  $\sigma_{_L} \times 10^5$    & G$_{_M}$/($\mu_p$ G$_{_D}$)   &$\mu_p$G$_{_E}$/G$_{_M}$   \\
(GeV/c)$^2$     & 	 &                          &  (OPE)                    & (OPE)                     \\ \hline 
 5.994    &   $167  \pm 4$~~     &  $ 7.1 \pm 4.6$ 	 & $1.000 \pm 0.011$~ & 0.75 $\pm$ 0.25  \\
 7.020    &   $104  \pm 3$~~     &  $ 9.3 \pm 5.3$	 & $0.967 \pm 0.015$~ & 1.18 $\pm$ 0.35  \\ 
 7.943    &   $71.0 \pm 2.7$~    &  $ 4.1 \pm 3.9$	 & $0.943 \pm 0.018$~ & 1.0 $\pm$ 0.5  \\
 8.994    &   $49.8 \pm 1.7$~    &  $ 0.7 \pm 3.0$	 & $0.934 \pm 0.016$~ & 0.5 $\pm$ 1.2 \\ 
 9.840    &   $36.9 \pm 2.4$~    &  $ 1.9 \pm 3.5$	 & $0.909 \pm 0.029$~ & 1.1 $\pm$ 1.0 \\ 
 12.249   &   $18.0 \pm 0.8$~    &  $ 1.2 \pm 1.8$	 & $0.858 \pm 0.019$~ & 1.3 $\pm$ 1.1  \\ 
 15.721  &   ~$8.6 \pm 0.5$     &  $-0.2 \pm 1.2$~~~& $0.840 \pm 0.025$~ &  (-0.9 $\pm$ 2.8)$~ $  
\end{tabular}
\end{ruledtabular}
\end{table} 

\begin{acknowledgments}
We thank the Jefferson Lab accelerator and Hall A technical staff for their outstanding support of this experiment, 
which was part of the first run group to take data at JLab after the accelerator upgrade.
We thank N.~Kivel for careful reading of the manuscript and his valuable suggestions.
Communications with P.N.~Kirk, S.~Rock, and A.~Sill about technical aspects of their experiments are very much appreciated.
This work was supported in part by 
the Science Committee of Republic of Armenia under grant 21AG-1C085, 
the Natural Sciences and Engineering Research Council of Canada (NSERC),
the UK Science and Technology Facilities Council (Grant nos. STFC 57071/1 and STFC 50727/1),
the U.S. National Science Foundation grant PHY-1508272, and the U.S. Department of Energy, Office of Science, 
Office of Nuclear Physics under contracts DE-AC02-05CH11231, DE-AC02-06CH11357, and DE-SC0016577, 
and DOE contract DE-AC05-06OR23177, under which JSA, LLC operates JLab.
\end{acknowledgments}

\bibliographystyle{apsrev4-1}
\bibliography{GMp12_PRL.bib}

\end{document}

%% file: commands.tex
\usepackage{scalerel}
\usepackage{tikz}
\usetikzlibrary{svg.path}

\definecolor{orcidlogocol}{HTML}{A6CE39}
\tikzset{
  orcidlogo/.pic={
    \fill[orcidlogocol] svg{M256,128c0,70.7-57.3,128-128,128C57.3,256,0,198.7,0,128C0,57.3,57.3,0,128,0C198.7,0,256,57.3,256,128z};
    \fill[white] svg{M86.3,186.2H70.9V79.1h15.4v48.4V186.2z}
                 svg{M108.9,79.1h41.6c39.6,0,57,28.3,57,53.6c0,27.5-21.5,53.6-56.8,53.6h-41.8V79.1z M124.3,172.4h24.5c34.9,0,42.9-26.5,42.9-39.7c0-21.5-13.7-39.7-43.7-39.7h-23.7V172.4z}
                 svg{M88.7,56.8c0,5.5-4.5,10.1-10.1,10.1c-5.6,0-10.1-4.6-10.1-10.1c0-5.6,4.5-10.1,10.1-10.1C84.2,46.7,88.7,51.3,88.7,56.8z};
  }
}

\newcommand\orcidicon[1]{\href{https://orcid.org/#1}{\mbox{\scalerel*{
\begin{tikzpicture}[yscale=-1,transform shape]
\pic{orcidlogo};
\end{tikzpicture}
}{|}}}}

\usepackage{hyperref}

\usepackage{graphicx}
\usepackage{dcolumn}
\usepackage{bm}
\usepackage{subfigure}
\usepackage{epstopdf}
\usepackage{hyperref}
\usepackage{color}
\usepackage[utf8]{inputenc}

\newcommand{\gep}{G$_{_E}$}
\newcommand{\gmp}{G$_{_M}$}
\newcommand{\gepsq}{${\rm G}_{_E}^2$}

\newcommand{\gevsq}{(GeV/c)$^2$}
\newcommand{\qsq}{Q$^2$}

%% file: collaboration.tex
%
\author{M.~E.~Christy}
\affiliation{Hampton University{,} Hampton, Virginia 23669, USA} 
\author{T.~Gautam}
\affiliation{Hampton University{,} Hampton, Virginia 23669, USA} 
\author{L.~Ou} 
\affiliation{Massachusetts Institute of Technology{,} Cambridge{,} Massachusetts 02139, USA}   
\author{B.~Schmookler} 
\affiliation{Massachusetts Institute of Technology{,} Cambridge{,} Massachusetts 02139, USA} 
\author{Y.~Wang}  
\affiliation{William and Mary{,} Williamsburg{,} Virginia 23185, USA}  
\author{D.~Adikaram}
\affiliation{Thomas Jefferson National Accelerator Facility{,} Newport News{,} Virginia 23606, USA}
\author{Z.~Ahmed}
\affiliation{University of Regina{,} Regina{,} SK{,} S4S 0A2 Canada}
\author{H.~Albataineh}
\affiliation{Texas A $\&$ M University{,} Kingsville{,} Texas 77843, USA}
\author{S.~F.~Ali}
\affiliation{Catholic University of America{,} Washington{,} District of Columbia 20064, USA} 
\author{B.~Aljawrneh}
\affiliation{North Carolina A\&T State University, Greensboro, North Carolina 27411, USA}
\affiliation{Al Zaytoonah University of Jordan, Amman 117733, Jordan}
% ...
